\documentclass[twocolumn,preprintnumbers,amsmath,amssymb]{revtex4}
\usepackage{graphicx}% Include figure files
\usepackage{dcolumn}% Align table columns on decimal point
\usepackage{bm}% bold math

\begin{document}

\title{Strong Aharonov-Bohm oscillations in GaAs two-dimensional holes}

\author{B.~Habib, E.~Tutuc, and M.~Shayegan}
%\email[Corresponding author, email address:
%]{bhabib@princeton.edu}

\affiliation{Department of Electrical Engineering, Princeton
University, Princeton, NJ 08544, USA}

\date{\today}

\begin{abstract}
We measured Aharonov-Bohm resistance oscillations in a shallow
two-dimensional GaAs hole ring structure, defined by local anodic
surface oxidation. The amplitude of the oscillations is about 10\%
of the ring resistance, the strongest seen in a hole system. In
addition we observe resistance oscillations as a function of front
gate bias at zero magnetic field. We discuss the results in light of
spin interference in the ring and possible applications to
spintronics.
\end{abstract}

\maketitle

%%%%%%%%%%%%%%%%%%%%%%%%%%%%%%%%%%%%%%%%%%%%%%%%%%%%%%%%%%%%%%%%%%
% \section{\label{sec:level1}Introduction}

The Aharonov-Bohm (AB) effect and related phenomenon in mesoscopic
semiconductor structures have attracted much attention over the last
years. The AB effect reflects the fact that when magnetic flux
pierces through a ring structure, a phase difference develops
between the electron wavefunctions traveling in the ring's two arms
\cite{AharonovPR59}. This phase difference is $2\pi(\Phi/\Phi_0)$,
where $\Phi = \pi r^2 B$ is the magnetic flux through the ring,
\emph{r} is the ring radius, $B$ is the applied magnetic field in
the perpendicular direction, and $\Phi_0 = h/e$ is the flux quantum.
Hence the resistance of the ring exhibits oscillations periodic in
$B$ with a period equal to $\pi r^2/(h/e)$.

In semiconductors, the AB effect has been seen in rings made from
two-dimensional (2D) electron systems in various materials
\cite{TimpPRL87GaoAPL94AppenzellerPRB95} but has been difficult to
observe in 2D hole systems (2DHSs). Recently, AB oscillations with
an amplitude of about 0.1\% (of the total ring resistance) were
observed by Yau \emph{et al.} \cite{YauPRL02} in a 500~nm radius
ring structure patterned via electron-beam lithography in GaAs 2D
holes. In the present work we aimed to increase the amplitude of the
AB oscillations in this system by fabricating smaller rings
\cite{footnote1}. For this purpose we employed the local anodic
oxidation (LAO) technique using an atomic force microscope (AFM).
Here we present our measurements of AB oscillations in a GaAs 2D
hole sample with $r\simeq 160$~nm. The observed AB oscillation
amplitude is about \emph{100 times} larger than in Ref.
\cite{YauPRL02}. A further motivation for our experiments is the
fact that the 2DHS in GaAs exhibits strong spin-orbit interaction
which is tunable with gate bias \cite{LuPRL98}. Nitta \emph{et al.}
\cite{NittaAPL99} highlighted that the resistance in a ring
structure in such a system can be modulated by changing the gate
bias even at $B=0$ and proposed it as a spin interference device. In
our ring structure, we do indeed observe oscillations as a function
of gate bias at $B=0$; however, the origin of these oscillations is
unclear.

%%%%%%%%%%%%%%%%%%%%%%%%%%%%%%%%%%%%%%%%%%%%%%%%%%%%%%%%%%%%%%%%%%
% \section{Experimental Results}

Our sample was grown on a GaAs (311)A substrate by molecular beam
epitaxy and contains a modulation-doped 2DHS confined to a
GaAs/Al$_{0.3}$Ga$_{0.7}$As interface. The interface is separated
from an 11~nm-thick Si-doped Al$_{0.3}$Ga$_{0.7}$As layer (Si
concentration of $1.64 \times 10^{19}$~cm$^{-3}$) by a 15~nm
Al$_{0.3}$Ga$_{0.7}$As spacer layer. A 5~nm GaAs layer caps the
structure resulting in the 2D hole layer residing $\simeq$~31~nm
below the surface. We fabricated Hall bar samples via optical
lithography and used alloyed In/Zn for the ohmic contacts. Metal
gates were deposited on the sample's front and back to control the
2D hole density ($p$). Before patterning the surface using an AFM
and depositing the gates, we characterized the 2DHS at $T \simeq$
300~mK and determined its mobility to be $1 \times 10^5$~cm$^2$/Vs
in the [$01\bar1$] direction at $p = 2.8 \times 10^{11}$~cm$^{-2}$.
The resistance measurements through the ring were made at $T \simeq$
30~mK in a dilution refrigerator with standard low frequency lock-in
technique.

We fabricated the ring structure via the LAO technique using an AFM.
A shallow 2D electron gas confined to a GaAs/AlGaAs heterostructure
is depleted when the surface of the sample is oxidized using a
negative bias on an AFM tip \cite{IshiiJJAP95}. The extent of the
depletion depends critically on the depth of the 2D gas and hence
the need to grow very shallow 2D samples as described above
\cite{HeldAPL98}. Recently, this technique was used to pattern
mesoscopic structures in GaAs 2DHSs \cite{RokhinsonSLM02,
GrbicAPL05, 2DEGrings}. Similar to these techniques, in order to
deplete the carriers in our structures, we used a tip voltage of
$-27.5$~V, a humidity level of $\sim 50$\% and a scanning speed of
$0.05~\mu$m/s.

\begin{figure*}
\centering
\includegraphics{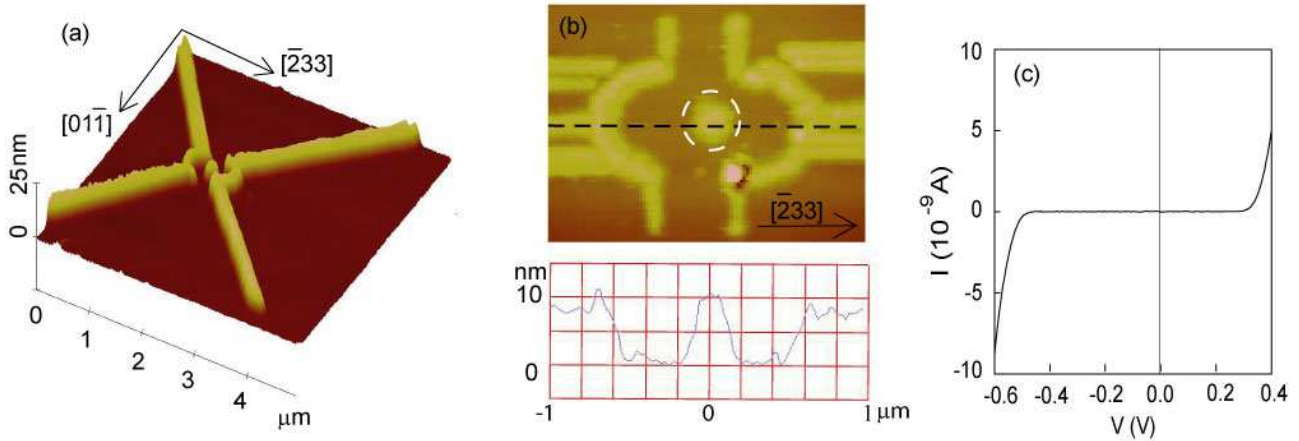}
\caption{\label{fig:ring}(Color on-line)(a) AFM micrograph of a
typical ring structure fabricated via the LAO technique. The light
(yellow) areas indicate the oxide. (b) AFM image of the ring used in
the experiment. The white dashed circle indicates the area of the
ring determined from the frequency of the AB oscillations. The
height profile along the black dashed line is shown in the bottom
graph. The oxidized ridges are $\sim10$~nm tall. (c) Current-voltage
profile across an oxidized ridge 10~nm in height, shows that the
ridge is insulating in the range of -0.5~V to 0.3~V.}
\end{figure*}

An AFM micrograph of one of our ring structures is shown in Fig.
\ref{fig:ring}(a). The light areas indicate the oxide grown with the
LAO technique. The drawn inner radius of the ring used for the
measurements is 100~nm [Fig. \ref{fig:ring}(b)]. With the oxidation
parameters stated above, we expect an oxide height of 10-12~nm. The
lower graph in Fig. \ref{fig:ring}(b) shows the height profile along
the dashed line in the top picture and indicates a well-formed,
10~nm-high profile. In order to demonstrate that the oxide lines do
indeed deplete the 2D gas underneath, we applied voltage (V) across
one side of such a line and measured the current leaking through the
barrier. There is negligible leakage for $-0.5 \lesssim$ V $\lesssim
0.3$~V [Fig. \ref{fig:ring}(c)].

\begin{figure*}
\centering
\includegraphics{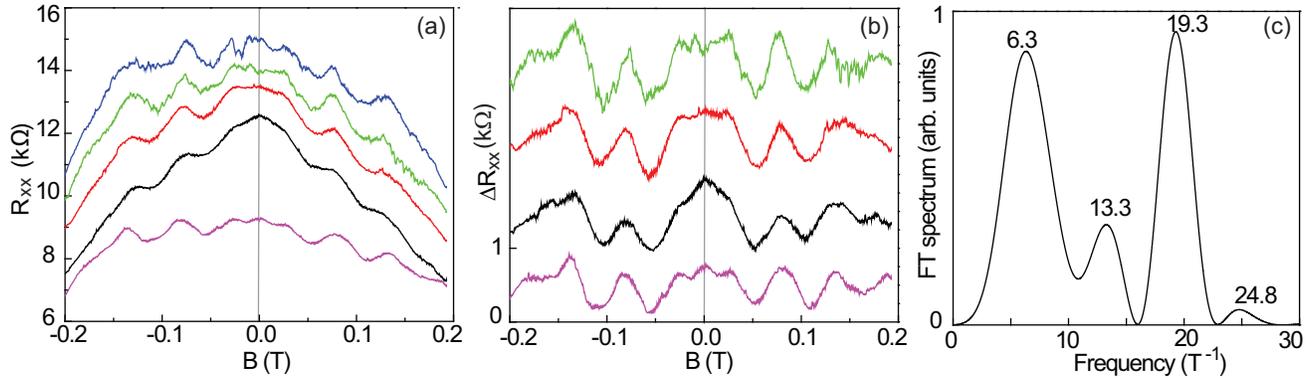}
\caption{\label{fig:osc}(Color on-line)(a) Measured AB oscillations
in the ring structure. The top four traces were measured at V$_B =
-140$~V and at V$_F = -12,-13,-14$ and -15~mV respectively, starting
from the top. The lowest trace was measured at V$_B = -143$~V and
V$_F = -5$~mV. The fourth trace from the top is shifted down by
$\sim1$~k$\Omega$ for clarity. (b) Same data as in (a) for bottom
four traces after subtracting the background resistance; traces are
shifted vertically for clarity. (c) The Fourier transform spectrum
of the measured oscillations, in the range $\pm0.2$~T, at V$_B =
-140$~V, V$_F = -13$~mV. The peak at 19.3~T$^{-1}$ corresponds to a
ring radius of 160~nm, consistent with the geometry of the ring (see
Fig. \ref{fig:ring}(b)).}
\end{figure*}

Figures \ref{fig:osc}(a) and (b) show the measured ring resistance
as a function of \emph{B} and the change in resistance after
subtracting a fourth order polynomial. The amplitude depends on the
front (V$_F$) and back (V$_B$) gate voltages, and the largest we
have seen is $\sim$10\% (at V$_F$ = -5~mV, V$_B$ = -143~V). In fact,
the resistance oscillations are very sensitive to the gate biases
and for a particular V$_B$, are only observed within a 2-3~mV range
of V$_F$. However, within this range the oscillations are very
robust and reproducible. The Fourier transform spectrum of one of
the traces, obtained by zero padding the data and Hamming window, is
shown in Fig. \ref{fig:osc} (c). The strong peak at 19.3~T$^{-1}$ is
observed in the Fourier transforms of all the traces and corresponds
to $r = 160$~nm which agrees with the drawn dimensions of the ring
[see dashed circle in Fig. \ref{fig:ring}(b)]. The origin of the
other peaks is less clear; the two side peaks at 13.3~T$^{-1}$ and
24.8~T$^{-1}$ may be related to spin-orbit coupling in the system
\cite{YauPRL02, CapozzaPRL05}. However, in our case, the number of
oscillations in the range of $\pm$0.2~T is small and it is difficult
to draw definitive conclusions.

We now discuss resistance oscillations in a ring device as a
function of V$_F$ at $B=0$. In a symmetric ring with spin-orbit
coupling, even in the absence of \emph{B}, the carrier wavefunction
acquires a phase difference when traversing the two arms of the ring
\cite{NittaAPL99}. This results in a constructive or destructive
interference at the other end of the ring. The phase difference is
proportional to $rk_d$, where $k_d$ is the difference of the Fermi
wavevectors ($k_\pm$) of the two spin-subbands which are split
because of the spin-orbit interaction. If we assume isotropic Fermi
contours, then $k_\pm = \sqrt{4\pi p_\pm}$ where $p_\pm$ are the
densities of the two spin-subbands. In 2D systems with spin-orbit
coupling, $k_d$ can be changed by tuning the electric field via
front and back gate biases \cite{LuPRL98, NittaPRL97}. Using this
fact, Nitta \emph{et al.} \cite{NittaAPL99} showed that the ring
resistance should oscillate as a function of the gate bias with the
period given by the condition, $r\Delta k_d\simeq 1$, where $\Delta
k_d$ is the change in $k_d$ with the gate bias \cite{footnote3}. A
ring structure in these systems can hence be used as a spin
interference device.

\begin{figure}
\centering
\includegraphics{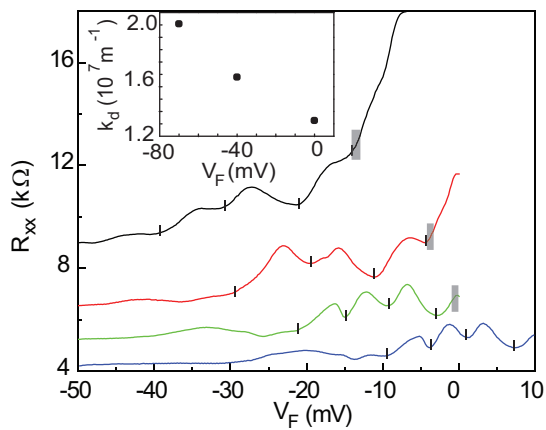}
\caption{\label{fig:Vosc}(Color on-line) Ring resistance measured at
zero magnetic field, as a function of V$_F$ for V$_B$ = -140, -143,
-146, and -149~V (from top to bottom). The grey boxes indicate the
values of V$_F$ at which AB oscillations are observed. No AB
oscillations were observed for V$_B$ = -149~V. Inset: The change in
$k_d$ as a function of V$_F$ in the unpatterned region.}
\end{figure}

Figure \ref{fig:Vosc} shows the resistance of our ring measured as a
function of V$_F$ at different V$_B$. The traces indeed exhibit
oscillations which are approximately periodic (e.g., with a period
$\simeq$ 7~mV for V$_B$ = -146~V). These oscillations were very
robust over many cool-downs of the sample.

A quantitative understanding of Fig. \ref{fig:Vosc} data, however,
poses a problem. In order to check whether $ r\Delta k_d \simeq 1$
for an oscillation period in $R_{xx}$ vs V$_F$, we measured
Shubnikov-de Haas (SdH) oscillations in an unpatterned region of the
sample for different V$_F$. From the Fourier spectra of the SdH
oscillations, we deduce $\Delta k_d$ (Fig. \ref{fig:Vosc} inset)
\cite{LuPRL98}. Based on this data, we expect $\Delta k_d \simeq
6\times10^5$~m$^{-1}$ for a 7~mV change in V$_F$. For $r = 160$~nm,
this gives $r\Delta k_d = 0.1 \ll 1$. We would like to point out
that in similar experiments, K\"{o}nig \emph{et al.}
\cite{KonigPRL06} measured the ring resistance as a function of
V$_F$ in a HgTe/HgCdTe quantum well sample and they too, find a very
small value of $r\Delta k_d \simeq 0.08$ for a full oscillation of
ring resistance. It should be noted that electrons in the conduction
band of HgTe are a spin-3/2 system, similar to holes in GaAs. The
spin orientation and spin precession of a spin-3/2 system are quite
subtle \cite{CulcerPRL06} and it is possible that it is this
complexity that leads to the anomalous oscillations.

In the discussion above we have ignored two aspects of the ring
device. First, the device is initially pinched off and a negative
V$_F$ has to be applied to populate the ring with holes. This
implies that the 2D hole density in the ring is much smaller than in
the unpatterned region. Second, it is apparent in Fig.
\ref{fig:ring}(b) that the ring is asymmetric. This asymmetry could
also lead to resistance oscillations at $B=0$ \cite{PedersenPRB00}.
The resistance modulation would then be a consequence of the
difference in the wavevectors in the right and left arms of the
ring, $k_{rl}$. If we assume an initial density in the range of
$1\times 10^{9}$~cm$^{-2}$ to $1\times 10^{10}$~cm$^{-2}$ in the
right arm and 5 times this density range in the left arm, a
plausible assumption given that the right arm is more pinched than
the left arm, we find $r\Delta k_{rl} \sim 0.7$, where $\Delta
k_{rl}$ is the change in $k_{rl}$ with V$_F$. Hence, low density and
asymmetry in the ring could possibly explain the oscillations
observed in our ring structure as a function of V$_F$
\cite{footnote4}. Note that this conjecture implies that the holes'
phase coherence is maintained in the V$_F$ range where the
oscillations in Fig. \ref{fig:Vosc} are seen. If so, it is then
puzzling why the AB oscillations are observed in a very narrow range
of V$_F$ (marked in Fig. \ref{fig:Vosc} by the grey boxes) for a
given value of V$_B$.

In conclusion, we measured AB oscillations in GaAs 2D holes in a
ring structure fabricated via the LAO technique. The observed
oscillations have the highest amplitude seen in this system and are
about \emph{100 times} larger than previously measured in rings
fabricated via electron-beam lithography. We also observe
oscillations with front gate bias whose origin remains a puzzle.

We thank K. Ensslin, B. Grbic and L. Rokhinson for helpful advice on
the AFM LAO technique, and R. Winkler and S. Lyon for discussions.
Our work was supported by the DOE, NSF and the von Humboldt
Foundation.

\end{document}